\documentclass[11pt,twoside]{article}

\usepackage{asp2014}

\aspSuppressVolSlug
\resetcounters

\begin{document}

\title{Non-negative matrix factorization approach to sky subtraction for optical spectroscopy}

\author{Fedor~Kolganov,$^1$ Igor~Chilingarian,$^2$ and Kirill~Grishin$^3$}
\affil{$^1$Leipzig University, Faculty of Physics and Earth System Sciences, Linnestr. 5, 04103 Leipzig, Germany; \email{fedor.kolganov@voxastro.org}}
\affil{$^2$Center for Astrophysics --- Harvard and Smithsonian, 60 Garden Street MS09, Cambridge, MA 02138, USA}
\affil{$^3$Universite Paris Cite, CNRS(/IN2P3), Astroparticule et Cosmologie, 10 rue Alice Domon et Leonie Duquet, 75013, Paris, France}

\paperauthor{Fedor~Kolganov}{kolganov@voxastro.org}{}{Universität Leipzig}{Author1 Department}{Leipzig}{State/Province}{Postal Code}{Germany}
\paperauthor{Igor~Chilingarian}{igor.chilingarian@cfa.harvard.edu}{ORCID_Or_Blank}{Author2 Institution}{Author2 Department}{City}{State/Province}{Postal Code}{Country}
\paperauthor{Kirill~Grishin}{Author3Email@email.edu}{ORCID_Or_Blank}{Author3 Institution}{Author3 Department}{City}{State/Province}{Postal Code}{Country}



\begin{abstract}
Numerous sky background subtraction techniques have been developed since the first implementations of computer-based reduction of spectra. \citet{2000ApJ...533L.183K} described a singular value decomposition-based method which allowed them to subtract night sky background from multi-fiber spectroscopic observations without any additional sky observations. We hereby take this approach one step further with usage of non-negative matrix factorization instead of principal component analysis and generalize it to 
2D spectra. This allows us to generate approximately 10 times as many valid eigenspectra because of non-negativity.
We apply our method to “short”-slit spectra of low-mass galaxies originating from intermediate-resolution Echelle spectrographs (ESI at Keck, MagE at Magellan, X-Shooter at the VLT) when sources fill the entire slit. We demonstrate its efficiency even when no offset sky observations were collected.
\end{abstract}



\section{Introduction and Motivation}
Subtraction of the sky background from the science exposures has always been a part of the reduction process of optical spectra. Over the years, various techniques were developed suitable for different types of spectra. If the extent of an object is smaller than the slit length, the sky background can be removed by interpolation [in a linearized spectrum] from the outer regions of the slit thus forming the approximation of the background spectrum, which is then subtracted from the original frame. A technique proposed by \citet{2003PASP..115..688K} is conceptually similar but it is applied to long-slit or multi-slit \citep{2015PASP..127..406C,2019PASP..131g5005K} spectra prior to resampling and possibly accounting for line-spread-function variations along the slit \citep{2011ASPC..442..143K}. If the slit is entirely filled by the science target, additional observation of offset sky background is usually utilized. Although it is a widely used technique, because the night sky spectrum changes rapidly in time, often much quicker than the typical duration of an exposure, the resulting offset sky spectrum cannot always precisely represent the background in the scientific observation. \cite{2000ApJ...533L.183K} were the first to introduce singular value decomposition (SVD) analysis of hundreds of sky spectra as an approach to model a single spectrum of the night sky in a given science exposure. Although their technique proved to be effective, it was developed for and applied to 1D spectra only and, therefore requires the removal of both sky and galactic continua, because in 1D they are degenerated and cannot be separated.

We propose a novel technique based on the 2D nature of spectra in case of long-slit observations. The additional spatial dimension allows one to decouple sky and galactic continua, since their profiles on the slit have different shape: the profile of the sky background is constant, while the source profile has a peak and/or gradient. Another novel aspect is the utilization of Non-Negative Matrix Factorisation (NNMF) to derive the eigenvectors for the sky model. We opted for it based on physical nature of sky spectral components: they are not expected to have negative flux components. 

\section{Description of the algorithm}
At the preparatory stage of the algorithm, the NNMF components need to be computed. The NNMF technique allows one to factorize an input matrix into two, with the constraint that all values of all three matrices remain non-negative:
$$
A \simeq W \cdot C,
$$
where A  is a matrix of $\text{N}_{\text{spec}}$ input sky spectra, $\text{N}_{\text{spec}} \times \text{N}_{\text{wl}}$; C is a matrix of components, $\text{N}_{\text{comp}} \times \text{N}_{\text{wl}}$ and W is a matrix of the weights for each component $\text{N}_{\text{spec}} \times \text{N}_{\text{comp}}$ (not needed for further calculations).

Currently, we make 3 general assumptions about the input source spectrum: (1) a sky spectrum does not vary along the slit; (2) ``non-continuum'' features of the science source (i.e. kinematic peculiarities of a galaxy, strong emission lines) do not occupy a significant fraction of the science spectrum in wavelength; (3) the photometric profile of a galaxy can be approximated by Moffat function \citep{1969A&A.....3..455M}. The algorithm can be divided into 2 major steps: (1) the galactic continuum removal; (2) a construction of the sky model based on the collapsed galaxy-removed spectrum and pre-computed NNMF components.

In the first step we assume that the flux in the sky spectrum is independent of the position on the slit. This can be achieved by a properly designed flat-fielding procedure and correction for the slit illumination. The flat profile of the sky will vanish when subtracted from itself. We first bin a 2D spectrum wavelength-wise in order to increase the signal-to-noise-ratio (SNR) with bin size not exceeding 30 pixels, since this step is aimed on a general approximation of the galactic continuum and does not require its precise reconstruction. In each bin the profile $F(y)$ consists of 2 components: `flat' sky and a galaxy:
$$
F(y) = \text{SKY} + \text{GAL}(y),
$$
where $y$ represents the spatial coordinate. Then we subtract a binned spectrum from itself, but offset along the slit by some value $\Delta y$:
$$
F(y + \Delta y) - F(y) =  [GAL(y + \Delta y) + SKY] - [GAL(y) + SKY] = GAL(y + \Delta y) - GAL(y),
$$
leaving us with the difference of the two galactic profiles,  which is then fitted by difference of two Moffat functions to obtain the parameters of the galactic profile in the current bin. From the values obtained in each bin, the approximation for the galactic continuum is constructed and then subtracted from the original frame. The bins that contain emission lines or strong kinematic peculiarities (e.g. in deep absorption lines), which cannot be modeled by a simple Moffat profile, are masked out.

Since the major spatial variations of the profile were removed from the spectrum in the Step 1, it is collapsed along the slit, increasing the pixel SNR. The resulting 1D spectrum is subject to modelling by NNMF components. The modelling is done by solving an overdefined system of linear equations for the vector $x$ of the component weights:
$$
s = C^T \cdot x + r,
$$
where $s$ is the residual sky spectrum from Step 1, collapsed along the slit and transposed; $C^T$ is transposed matrix of the NNMF components; $x$ are weights to be found; $r$ is a vector of residuals, which undergoes the least-squares minimization. After the weights $x$ have been found, a constructed sky model is subtracted from the original 2D spectrum leaving one with the spectrum of the science source.


\begin{figure}[!htb]
   \begin{minipage}[t]{0.499\textwidth}
     \centering
     \includegraphics[width=\linewidth]{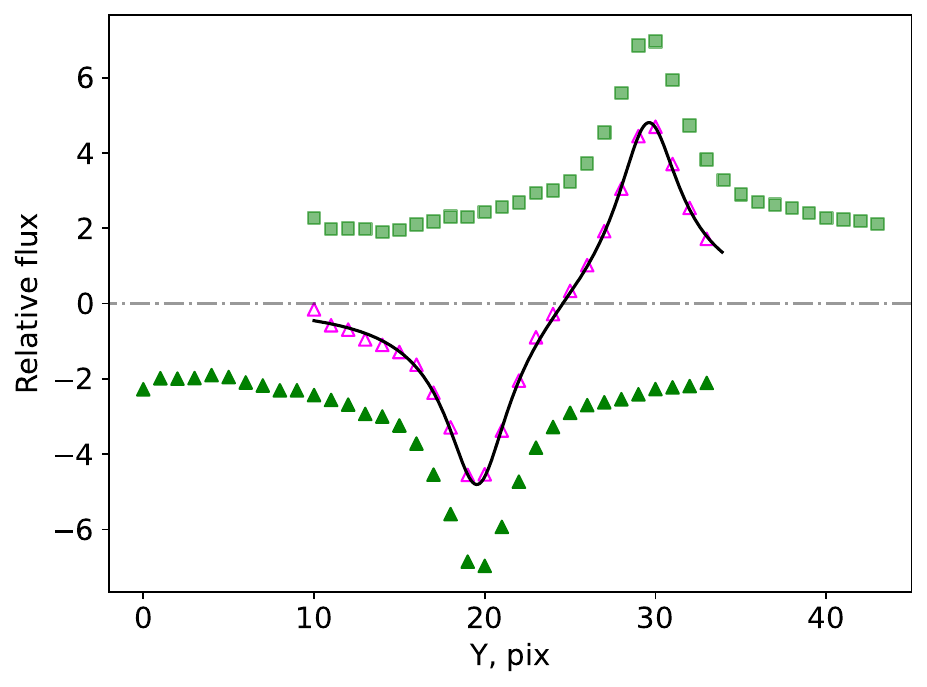}
   \end{minipage}\hfill
   \begin{minipage}[t]{0.499\textwidth}
     \centering
     \includegraphics[width=\linewidth]{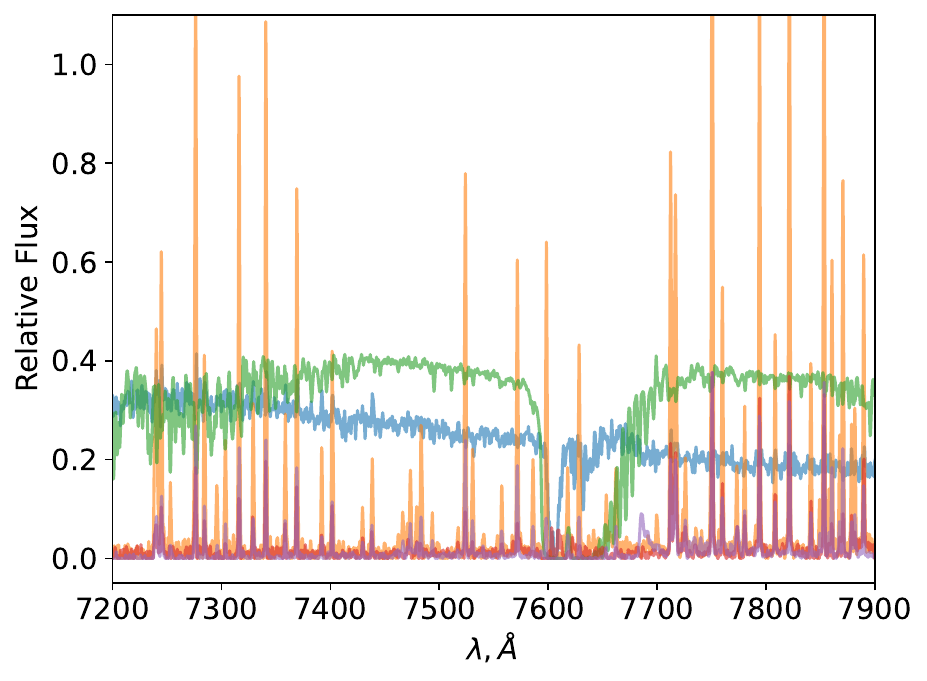}
   \end{minipage}
   \caption{\textbf{Left:} An illustration to our procedure to obtain the parameters of a Moffat profile in a single wawelength bin. \textit{Green triangles} show the inverted total profile; \textit{Light green squares} show the total shifted profile; \textit{Magenta triangles} display the difference; \textit{Black line} represents a fit with the difference of 2 Moffat profiles. \textbf{Right:} Subsection of the first 5 NNMF components from a sample of 150 sky spectra from $7200 \AA$ to $7900 \AA$.}\label{fig:prof_comp}
\end{figure}

\begin{figure}[!h]
    \centering
    \includegraphics[width=\textwidth]{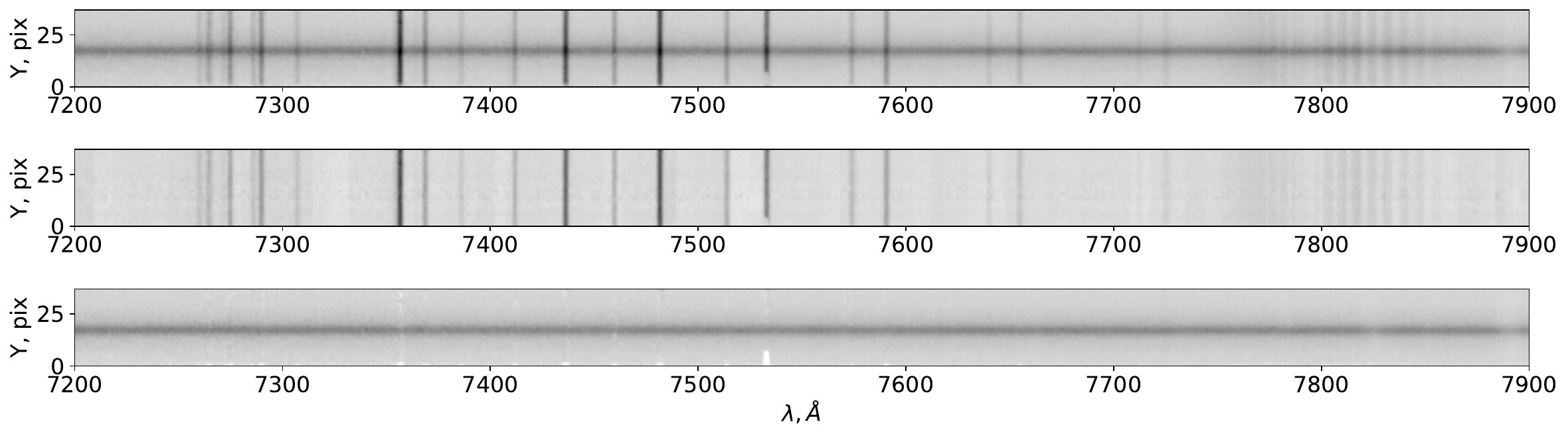}
    \caption{The same part of a galaxy spectrum at different steps of the workflow. \textbf{Top:} A subsection of the original MagE spectrum from $7200 \AA$ to $7900 \AA$. \textbf{Middle:} The same after the removal of the galactic continuum. \textbf{Bottom:} The original spectrum after subtraction of the sky background represented by 20 NNMF components.}
    \label{fig:res}
\end{figure}

\section{Performance of the algorithm on intermediate-resolution Echelle spectra}

We tested the algorithm on intermediate-resolution Echelle spectra taken with MagE spectrograph \citep{2008SPIE.7014E..54M} at the 6.5-m Clay Magellan II telescope at the Las Campanas Observatory, Chile. The NNMF sample consisted of around 200 sky spectra, from which we obtained 20 NNMF components. The sample contained a number of continuum-dominated spectra, which can be seen in the green and blue components in the right panel of Fig. \ref{fig:prof_comp} We then performed the sky subtraction on a spectrum of a galaxy with dimensions larger than the length of the slit. The stages of the workflow can be seen in Fig. \ref{fig:res}.

\section{Summary}
We present the first version of the novel sky subtraction technique for optical spectroscopy. It allows one to perform background subtraction for 2D spectra without consecutive observations of the offset-sky spectrum. In the upcoming validation process we are going to (i) increase sample of the sky spectra, which are decomposed into eigenvalues, using data archives of the instruments; (ii) apply the technique to Echelle spectra from ESI \citep{2002PASP..114..851S}, MagE, and X-Shooter \citep{2011A&A...536A.105V}; (iii) combine it with an approach from \citet{2003PASP..115..688K} in order to remove interpolation noise in the sky emission lines.

\bibliography{C104}  


\end{document}